\documentclass[a4paper,12pt]{article}

\usepackage{subfigure}
\usepackage{setspace}
\usepackage{multirow} 
\usepackage{multicol}
\usepackage{xspace}
\usepackage{amsbsy} 
\usepackage{graphics} 
\usepackage{color}
\graphicspath{./}

\usepackage{rotating}
\usepackage[version=3]{mhchem}
\usepackage[left=2.5cm,right=2.5cm,top=2cm,bottom=2cm]{geometry}

\usepackage{fancyhdr}
\fancypagestyle{plain}{ %
\fancyhf{} 
\lhead{\textit{ORs between icosahedral clusters}}
\rhead{J. M. Rosalie, A. Singh, H. Somekawa \& T. Mukai}
\lfoot{\textit{Corrected proof}}
\rfoot{\thepage}
}

\newcommand{\betap}{\ensuremath{{\beta_1^\prime}}\xspace}
\newcommand{\betapp}{\ensuremath{{\beta_2^\prime}}\xspace}
\newcommand{\iphasecomp}{\ensuremath{\mathrm{Mg_3Zn_6Y}}\xspace}
\newcommand{\micron}{\ensuremath{\mathrm{\mu m}}\xspace}
\newcommand{\degree}{\ensuremath{\mathrm{^o}}\xspace} 
\newcommand{\celsius}{\ensuremath{\mathrm{^oC}}\xspace} 

\newcommand{\fig}[2][6cm]{\resizebox{#1}{!}{\includegraphics{#2}}} 

\begin{document}
\title{Orientation relationships between icosahedral clusters in hexagonal  {\ce{MgZn2}} and monoclinic {\ce{Mg4Zn7}} phases in Mg-Zn(-Y) alloys} 

\author{Julian~M~Rosalie$\mathrm{^{a}}$ \and Hidetoshi Somekawa$\mathrm{^{a}}$ \and Alok Singh$\mathrm{^{a}}$,
 \and Toshiji Mukai$\mathrm{^{a}}$}
\date{}

\maketitle
\noindent
$\mathrm{^{a}}$Lightweight Alloys Group, Structural Metals Center, 
National Institute for Materials Science (NIMS), Sengen 1-2-1, Tsukuba,
Ibaraki, 305-0047, Japan.
\vspace{6pt}

\begin{abstract}
Intermetallic precipitates formed  in heat-treated and aged Mg-Zn and Mg-Zn-Y alloys have been investigated via electron microscopy. 
Coarse spheroidal precipitates formed on deformation twin boundaries contained domains belonging to either the \ce{MgZn2}~hexagonal Laves phase or the monoclinic \ce{Mg4Zn7}~phase. 
Both phases are structurally related to the quasi-crystalline phase formed in Mg-Zn-Y alloys, containing icosahedrally coordinated zinc atoms arranged as a series of broad rhombohedral units. 
This rhombohedral arrangement was also visible in intragranular precipitates where local regions with the structures of hexagonal \ce{MgZn2} and \ce{Mg4Zn7} were found. 
The orientation adopted by the  \ce{MgZn2} ~and \ce{Mg4Zn7}~phases in twin-boundary and intragranular precipitates was such that the icosahedral clusters were aligned similarly. 
These results highlight the close structural similarities between the precipitates of the Mg-Zn-Y alloy system.
\end{abstract}


\section*{Introduction}

Magnesium-zinc-yttrium alloys with Zn:Y ratios $\sim$6 precipitate an icosahedral quasi-crystalline phase (i-phase) with composition  \iphasecomp \cite{Luo1993,Tsai1994}. 
The binary Mg-Zn phases formed in these alloys play a key role in the precipitation strengthening of both commercial (ZK series) and experimental magnesium alloys. 
Aside from their industrial importance, two of these phases, monoclinic \ce{Mg4Zn7} and hexagonal  \ce{MgZn2}, contain zinc-centred icosahedral atomic clusters. 
This structural similarity to the i-phase has lead to these phases being used as structural models for the quasi-crystalline phase  \cite{Luo1993}.  

The \ce{Mg4Zn7} phase (a=2.596\,nm, b=0.524\,nm, c=1.428\,nm, \(\beta =102.5\degree\) \cite{Yarmolyuk1975} )
has been shown to be structurally related to the i-phase \cite{Yang1987} and to the decagonal phase in Mg-Zn-Y alloys \cite{Yi2002} and can co-exist with the decagonal phase with the orientation relationship \(10f\parallel[010] D2a(G)\parallel[-401]\) ; and \(D2b(F)\parallel [102]\). 
In Mg-Zn alloys the \ce{Mg4Zn7} phase adopts an axial orientation   
\( [010]_\ce{Mg4Zn7} \parallel [0001]_{\mathrm{Mg}}\). 
Matching between precipitate and matrix lattice planes is more complex, 
with reports of  \((01\overline{1}0)_{\mathrm{Mg}}\) planes being aligned with 
  the \((200)_\ce{Mg4Zn7}\), \((202)_\ce{Mg4Zn7}\)  or \((603)_\ce{Mg4Zn7}\) planes \cite{Gao2007,Singh2007}.  
The relationship \( (\overline{2}110)_{\mathrm{Mg}}  \parallel (\overline{2}01)_\ce{Mg4Zn7} \) has also been reported \cite{Takahashi1973, Singh2007}.  
It has recently been suggested that two orientation relationships; \( [010]_\ce{Mg4Zn7} \parallel [0001]_{\mathrm{Mg}}\) with  
\((802)_\ce{Mg4Zn7} \parallel (01\overline{1}0)_{\mathrm{Mg}}\)   or 
\((603)_\ce{Mg4Zn7} \parallel (01\overline{1}0)_{\mathrm{Mg}}\)  may be sufficient to describe the relationship between the precipitate and matrix \cite{SinghPhilMag2010}. 

The \ce{MgZn2} phase (a=0.52\,nm, c = 0.85\,nm \cite{Sturkey1959} has a  \(C14\) hexagonal Laves phase structure and generally forms as plate-shaped precipitates on extended ageing. 
In this form it adopts an orientation relationship \( [0001]_{\mathrm{Mg}}\parallel [0001]_{\ce{MgZn2}} \);
\( (11\overline{2}0)_{\mathrm{Mg}}\parallel (10\overline{1}0)_{\mathrm{MgZn_2}} \) 
\cite{Takahashi1973,SinghTMS2008}. 
Less commonly, it has been reported to form as laths with \( [0001]_\mathrm{Mg}\parallel [11\overline{2}0]_{\ce{MgZn2}} \); 
\( (11\overline{2}0)_\mathrm{Mg}\parallel (0001)_{\ce{MgZn2}} \) \cite{Gao2007} .

A recent work has reported that a C15 cubic Laves phase can also be found inside \betap precipitates along with the previously reported hexagonal (C14) structure. 
The transformation between the two phases was explained in terms of a synchroshear mechanism whereby the stacking  sequence of several layers within the hexagonal phase underwent simultaneous displacements in the basal plane \cite{KimLaves2010}. 

It has also been shown that  the \ce{Mg4Zn7} and \ce{MgZn2} phases can co-exist in Mg-Zn-Y alloys in both rod-shaped precipitates (commonly termed \betap) \cite{SinghPhilMag2010} and larger, roughly spheroidal twin-boundary precipitates  \cite{RosaliePhilMag2010}. 
The co-existence of these phases within individual precipitates reconciles conflicting reports on rod-shaped precipitates in Mg-Zn(-Y) alloys which have been described as either  \ce{MgZn2} \cite{Sturkey1959,Takahashi1973, Mendis2009} or \ce{Mg4Zn7} \cite{Gao2007,Singh2007}. 
The \ce{MgZn2}-\ce{Mg4Zn7} interface was described as a largely continuous patterning of rhombohedral units whose corners corresponded to the location of the icosahedrally coordinated zinc atoms \cite{RosaliePhilMag2010}. 
This indicates that the locations of the centres of the clusters remain consistent across the interface, 
but did not address the  issue of the location of  the surrounding atoms.  
These studies were conducted on Mg-Zn-Y alloys and while yttrium has not been detected in either rod or shaped precipitates \cite{Singh2007,WeiPrec1995},  it was not possible to conclusively rule out yttrium playing a role in altering the structures of \ce{Mg4Zn7} or \ce{MgZn2}. 
The present work has examined the orientation of the two co-existing phases in both Mg-Zn and Mg-Zn-Y alloys to determine whether the orientation of the icosahedra changes across the interface between \ce{MgZn2} and \ce{Mg4Zn7}. 

\section*{Experimental details}

Magnesium alloys containing a) 3.0at.\%Zn and b) 3.0at.\%Zn and 0.5at.\%Y were produced via direct chill casting. 
The cast billet were homogenised for 15\,h at 300 and 350\celsius, respectively and extruded with extrusion ratio 12:1 at 300\celsius. 
Cylindrical samples machined from the rod were encapsulated in helium, solution treated (400 or 300 \celsius, respectively, for 1\,h) and quenched in water. 
These samples were either aged directly or alternatively compressed parallel to the extrusion direction to 3\% plastic strain at a strain rate of \(1\times10^{-3}\,\mbox{s}^{-1}\) prior to ageing.
This process of extrusion and compression served to generate deformation twins in the magnesium matrix which acted as additional nucleation sites.
Twins were extremely rare ($<1\%$ by volume) in equivalent samples not subjected to compressive deformation. 
Ageing was conducted at 150\celsius in an oil bath for periods from 8\,h to 216\,h. 

Foils for TEM and HRTEM examination were prepared from samples prepared as above, ground to \(\sim70\,\micron\)  thickness and finally thinned to perforation using a Gatan precision ion polishing system. 
A JEOL 4000EX instrument operating at 400\,kV was used for all high-resolution transmission electron microscopy in this work.  
HRTEM images were compared with multi-slice simulations performed on JEMS software \cite{StadelmannJems1987} using pre-loaded parameters for the JEOL 4000EX instrument. 
Structural models were constructed using VESTA software \cite{Vesta2008}.

\section*{Results}

Transmission electron microscope (TEM) images showed a microstructure consisting of rod-like \betap precipitates, aligned along the hexagonal axis of magnesium in foils of both Mg-Zn and Mg-Zn-Y alloys. 
For samples deformed in compression to generate twins, larger, roughly spheroidal precipitates were present at the twin boundaries.  
Plate-like precipitates with (0001) habit (termed \betapp) were observed on occasion, particularly in samples aged from longer periods. 
Ternary alloys also contained spheroidal intragranular precipitates, whose appearance was consistent with that of the i-phase. 

\begin{figure}
	\begin{center}
		\includegraphics[width=3.5in]{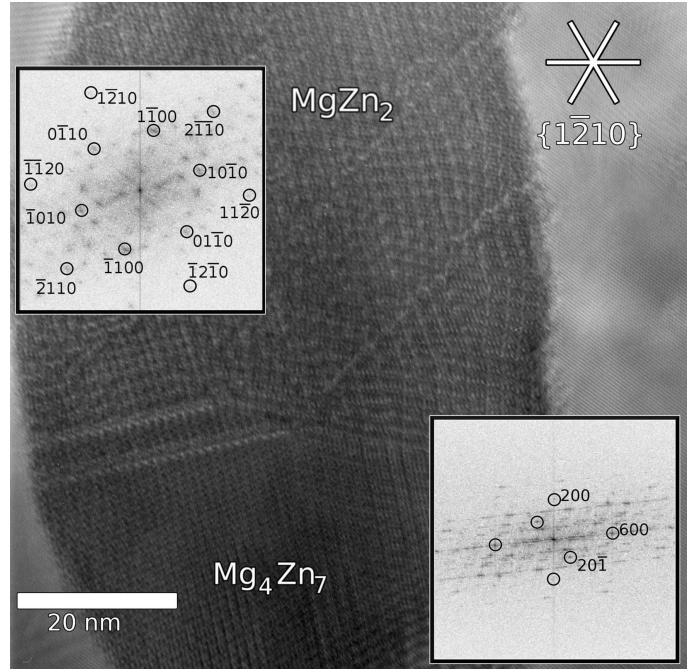}
		\caption{A precipitate formed on a twin boundary in Mg-3\%Zn showing domains of the \ce{MgZn2} (upper region) and 
\ce{Mg4Zn7} (lower region) phases.
Fast Fourier transforms of each phase are inset. 
The electron beam is aligned along the [0001] axis of the matrix (upper right potion of the field). 
The \ce{MgZn2} phase has orientation relationship
 \( (11\overline{2}0)_{\mathrm{Mg}}\parallel (10\overline{1}0)_{\mathrm{MgZn_2}} \) 
to the matrix, while the 
\ce{Mg4Zn7} phase has OR \( [010]_{\betap} \parallel [0001]_{\mathrm{Mg}}; (201)_{\betap} \parallel (1\overline{1}00)_\mathrm{Mg}\). 
\label{fig-hrtem-1}}
	\end{center}
\end{figure}

The precipitates on twin boundaries were divided into domains with structures of either \ce{MgZn2} or \ce{Mg4Zn7} phase.
Figure~\ref{fig-hrtem-1} shows a micrograph of a precipitate on a twin boundary in Mg-3at.\%Zn. 
Fast Fourier transforms (FFT) of the  upper portion of the precipitate show a structure corresponding to \ce{MgZn2}, with orientation  \(  [0001]_{\ce{MgZn2}} \parallel [0001]_{\mathrm{Mg}};   (11\overline{2}0)_{\mathrm{Mg}}\parallel (10\overline{1}0)_{\mathrm{MgZn_2}} \)  to the matrix.  
This region is subdivided into a number of domains. 
Fourier transforms of the lower portion of the figure show a structure resembling that of \ce{Mg4Zn7} with OR  
\( [010]_{\betap} \parallel [0001]_{\mathrm{Mg}}; (20\overline{1})_\ce{Mg4Zn7} \parallel (11\overline{2}0)_\mathrm{Mg}\). 
This arrangement results in a close match between the 
\( (0\overline{1}10)_\mathrm{MgZn_2 }\)
and
\( (20\overline{1})_\ce{Mg4Zn7}\) planes.

\begin{figure}
	\begin{center}
		\subfigure[\label{fig-201a}]{\fig[0.6\textwidth]{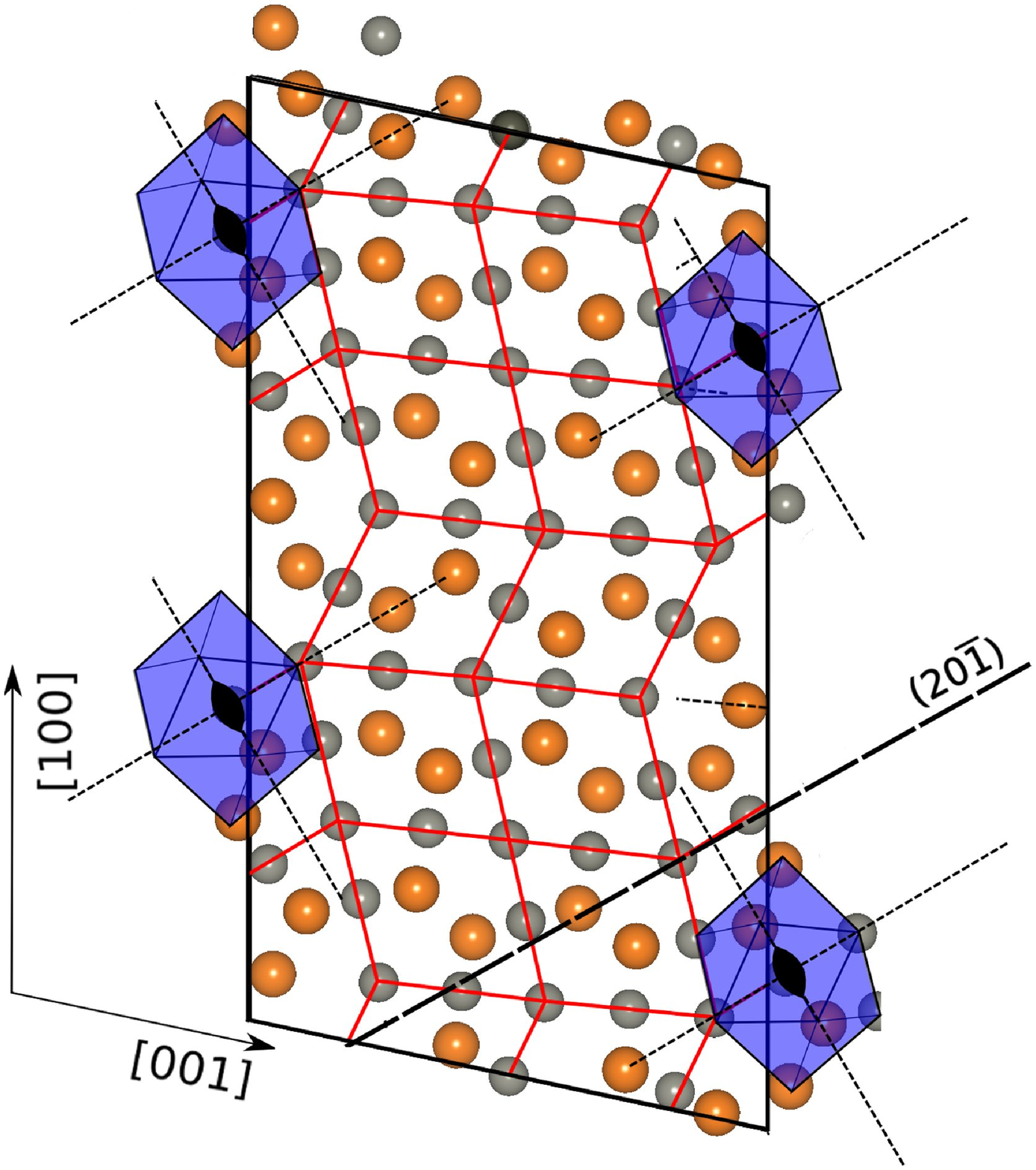}}
		\subfigure[\label{fig-201b}]{\fig[0.3\textwidth]{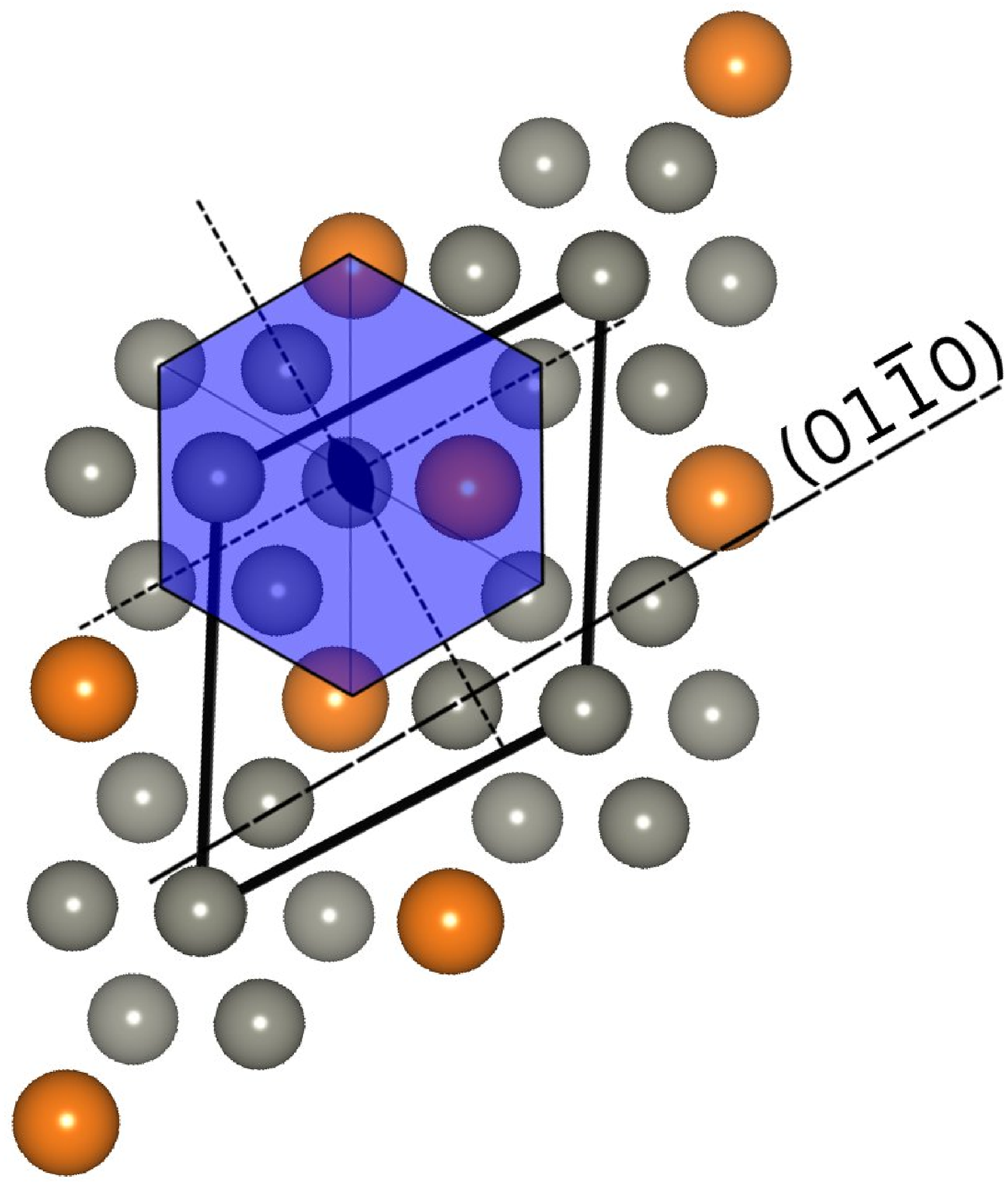}}
		\caption{Structural models of a) the \ce{Mg4Zn7} monoclinic phase and b) the hexagonal  \ce{MgZn2} Laves phase aligned as in Figure~\protect\ref{fig-hrtem-1}. 
a) shows the \ce{Mg4Zn7} structure with $(20\overline{1})$ plane indicated by a dashed line
b) shows the \ce{MgZn2} structure with the $(01\overline{1}0)$ plane indicated by a dashed line.
Icosahedral clusters in each structure are aligned with 2-fold axis i) normal to plane of the figure and ii) along the indicated planes.  
Icosahedral clusters in the monoclinic phase are distorted due to the atomic size effect, with Mg being larger than Zn. 
\label{fig-201}}
	\end{center}
\end{figure}

Structural models of the two phases for this orientation relationship are shown in Figure~\ref{fig-201}.
The two phases have been draw with \([010]_\ce{Mg4Zn7} \parallel [0001]_\ce{MgZn2}\)  and the 
 \((01\overline{1}0)_\ce{MgZn2}\) plane parallel to $(20\overline{1})_\ce{Mg4Zn7}$ as for the precipitate in Figure~\ref{fig-hrtem-1}.
Unit cell boundaries are shown in dark type, with the rhombohedral arrangement of icosahedral clusters in the monoclinic phase also indicated. 
Figure~\ref{fig-201a}(shows the \ce{Mg4Zn7} structure with the $(20\overline{1})$ plane indicated by a thick dashed line. 
Zinc atoms close to the edges of the unit cell have co-ordination number 12 \cite{Yarmolyuk1975} and define the centres of  icosahedral clusters, as indicated in the figure with their two-fold axes shown by dashed lines. 
These icosahedra are oriented with one 2-fold axis parallel to 	
\([010]_\ce{Mg4Zn7}\) and a second 2-fold axis lying in the \((20\overline{1})_\ce{Mg4Zn7}\) plane. 
Figure~\ref{fig-201b}  shows the hexagonal Laves phase structure where can be seen that the icosahedra in both phases share the same orientation relationship with the matrix; with one of the cube directions parallel to the hexagonal axis of magnesium and a second, perpendicular 2-fold axis is aligned normal to the
\( (20\overline{1})_\mathrm{Mg4Zn7} \) and 
\( (01\overline{1}0)_\mathrm{MgZn} \) planes.
The orientation relationship for these clusters can be described as;
\(  [0/0~0/0~/0/2] \parallel [0001]_\mathrm{Mg}\);  \(  [0/0~0/2~0/0] \parallel (11\overline{2}0)_\mathrm{Mg}\) with Cahn  indexing.

\begin{figure}
	\begin{center}
		\subfigure[]{\fig[10cm]{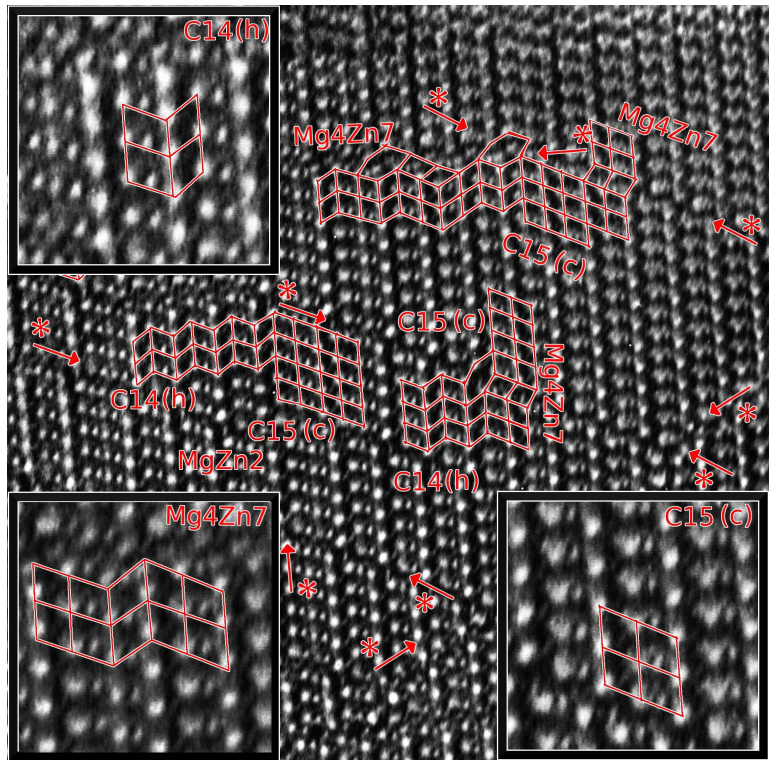}} \hfill

		\caption{\label{fig-hr-1}A high-resolution TEM micrograph showing a large intragranular precipitate in a Mg-3.0at\%Zn -0.5at.\%  alloy aged for 48\.h at 150$^o$C. 
The precipitate is viewed with the electron beam aligned parallel to \([0001]_\mathrm{Mg}\). 
Zinc-sites corresponding to the centres of icosahedral clusters show bright contrast and can be seen to be arranged as a series of edge-sharing rhombohedra. 
There is considerable disorder within the precipitate and several defects which appear to be linear in projection are indicated by asterisked arrows. 
These gives rise to regions of rhombohedra with the monoclinic \ce{Mg4Zn7} structure straddling the defect.
Between these defects the structure is that of the hexagonal \ce{MgZn2} Laves phase, with some regions which may correspond to a cubic Laves phase of the same composition. 
Regions of each structure are indicated in the figure.
Enlarged insets show regions corresponding to the  C14 hexagonal Laves phase (top left),  \ce{Mg4Zn7} monoclinic (bottom left) and the proposed $C15$ cubic Laves phase \protect\cite{KimLaves2010} (bottom right). }
	\end{center}
\end{figure}

The details of the structure could be more easily resolved in the intragranular precipitates in thin regions of the foil. 
Figure~\ref{fig-hr-1} presents a high-resolution TEM micrograph of a larger intragranular \betap precipitate in a foil of Mg-3.0at\%Zn -0.5at.\% alloy aged for 48\.h at 150$^o$C with the precipitates viewed in cross-section along the [0001] axis of magnesium. 
The  defocus condition is such that the icosahedrally coordinated zinc atoms show strong contrast and their arrangement into rhombohedral units can be seen. 
Despite the presence of defects, this arrangement was largely continuous throughout well-resolved regions of the precipitate.
The precipitate includes regions were the structure can be described as the \ce{MgZn2} C14 hexagonal Laves phase and \ce{Mg4Zn7} monoclinic phase.  
These regions are separated by defects (indicated by asterisked arrows) where the stacking sequence and arrangement of rhombohedra changes.
Defects were noted parallel to the (0001), \( (10\overline1{0} \) and  $(20\overline{2}3)$ planes of the hexagonal Laves phase. 
Some regions appeared consistent with presence of a \ce{MgZn2} C15 cubic Laves phase reported by Kim \textit{et. al}  \cite{KimLaves2010}, however, in the absence of images along several additional zone axes it was not possible to definitively identify this phase.
 
The continuity of the network of rhombohedra throughout the two structure indicates that the atomic positions of the icosahedral zinc clusters and the orientation of the rhombohedra were the same in both phases.

\begin{figure}
	\begin{center}
	\hfill Experimental \hfill Simulated \hfill\ 

	\hfill
	\begin{sideways}
	MgZn$_2$ C14 hex.
	\end{sideways}
	\subfigure[]{\includegraphics[width=0.3\textwidth]{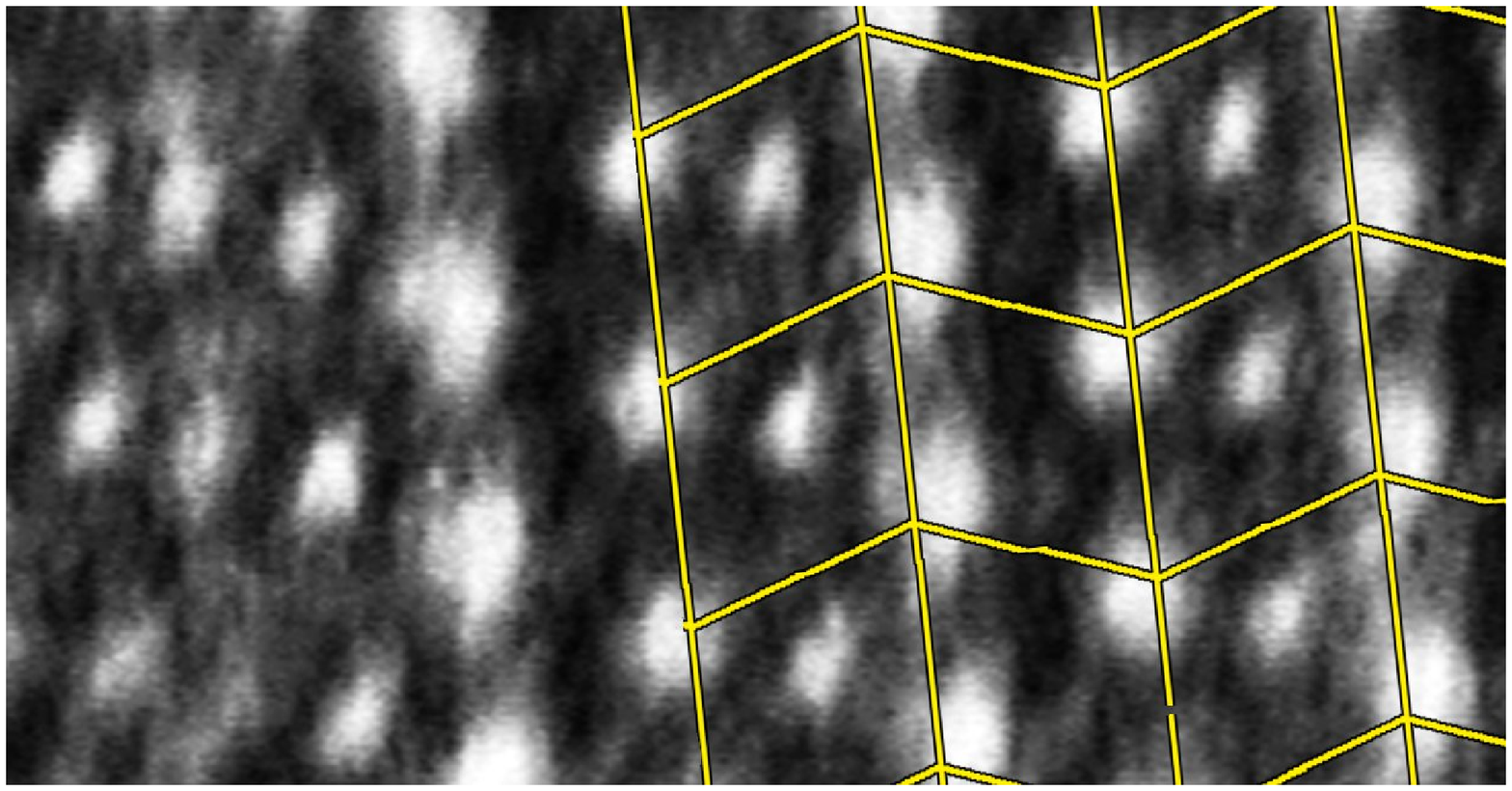}}\hfill
	\subfigure[]{\includegraphics[width=0.3\textwidth]{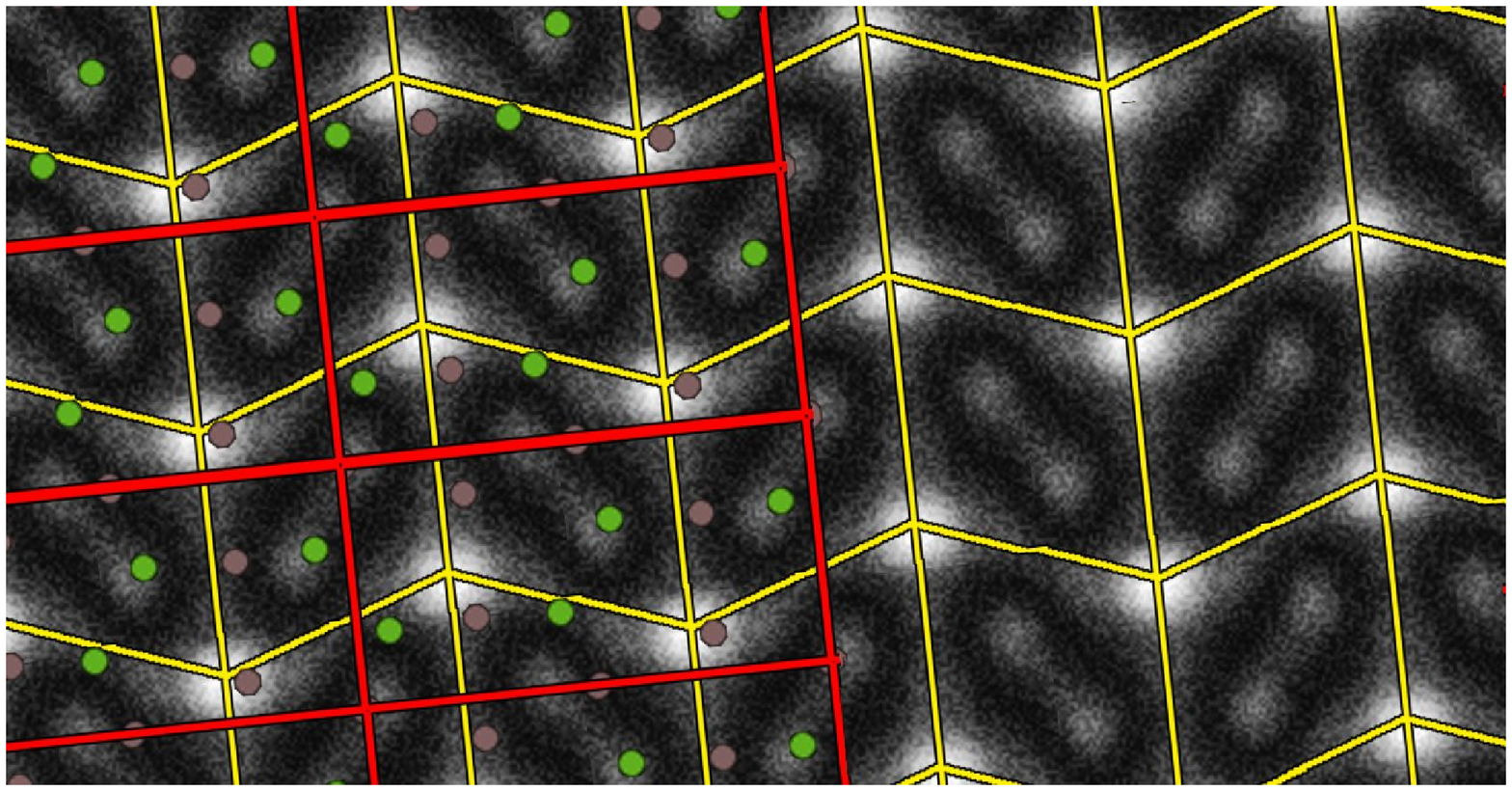}}\hfill\

	\hfill
	\begin{sideways}
	MgZn$_2$ C15 cubic.
        \end{sideways}
	\subfigure[]{\includegraphics[width=0.3\textwidth]{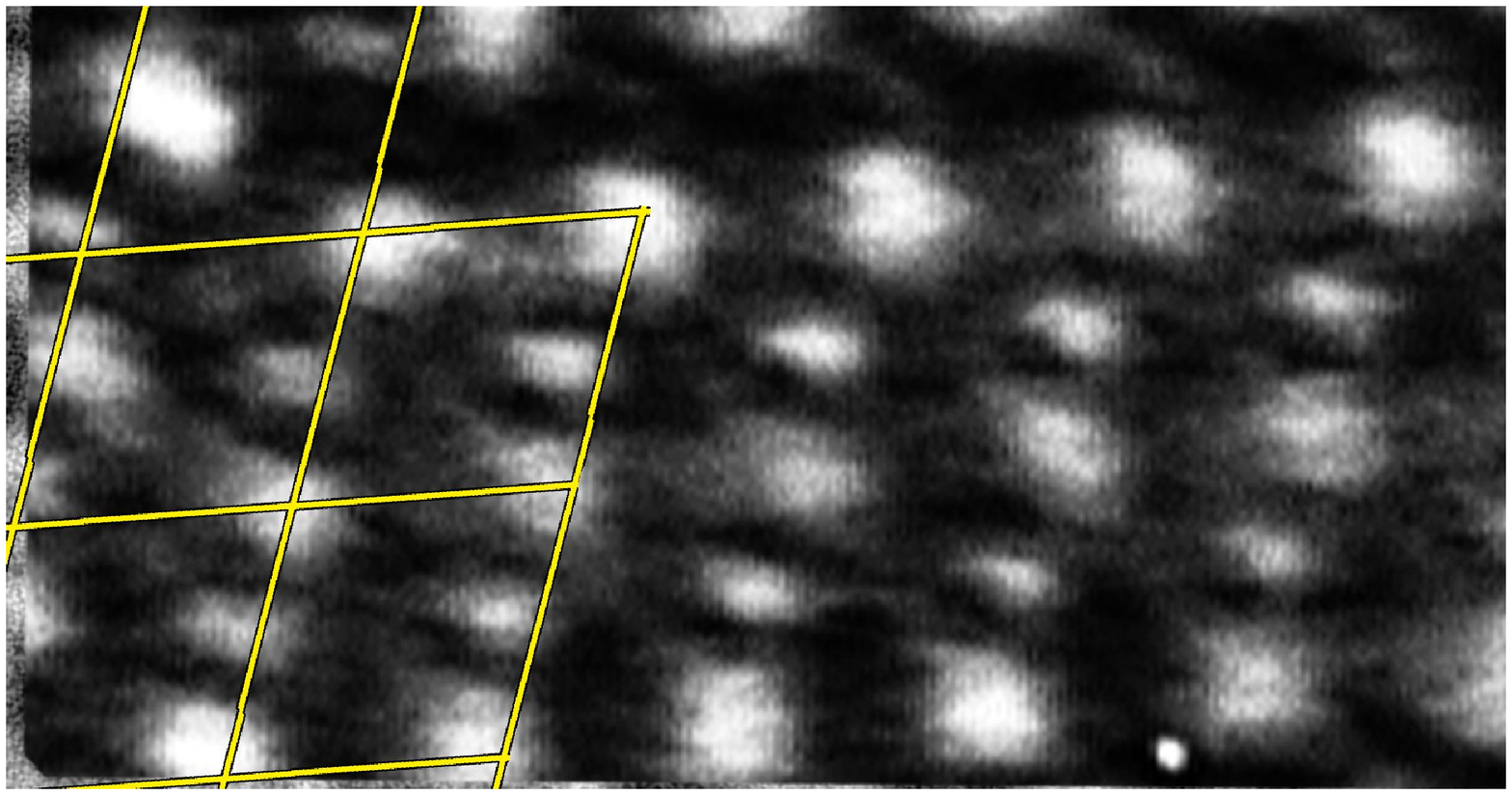}}\hfill
	\subfigure[]{\includegraphics[width=0.3\textwidth]{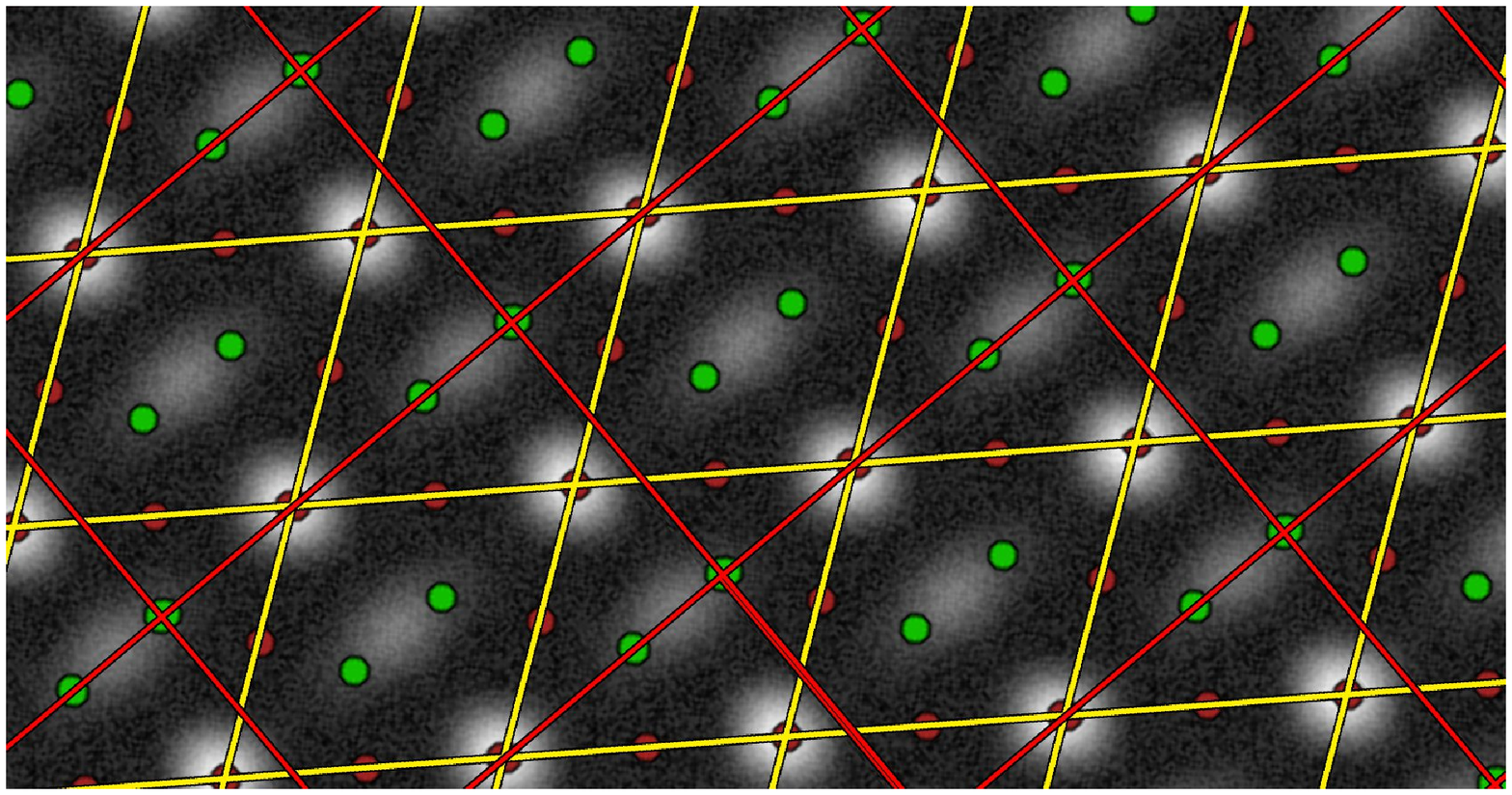}}\hfill\

	\hfill
	\begin{sideways}
	Mg$_4$Zn$_7$ Monoclinic
	\end{sideways}
	\subfigure[]{\includegraphics[width=0.3\textwidth]{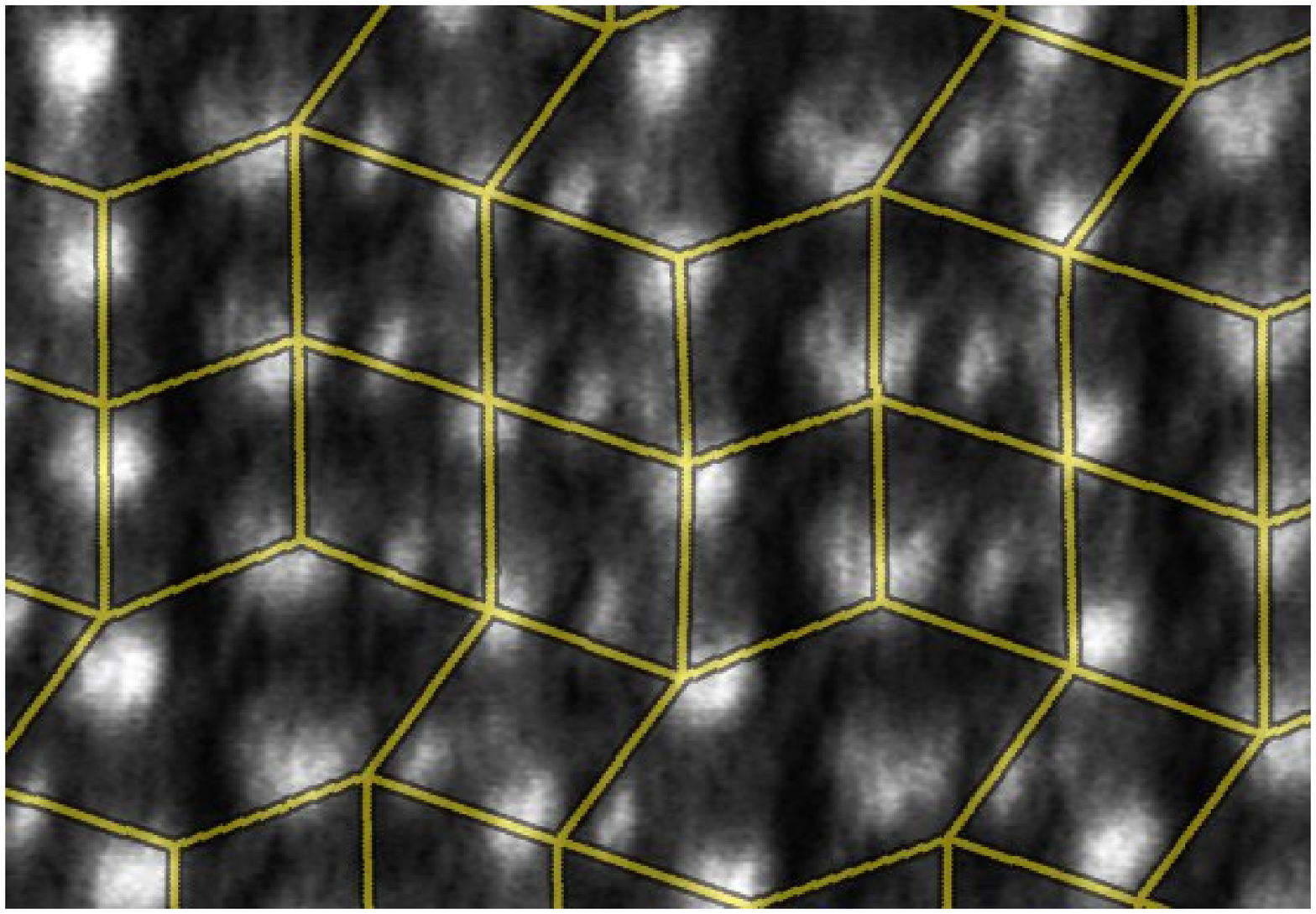}}\hfill
	\subfigure[]{\includegraphics[width=0.3\textwidth]{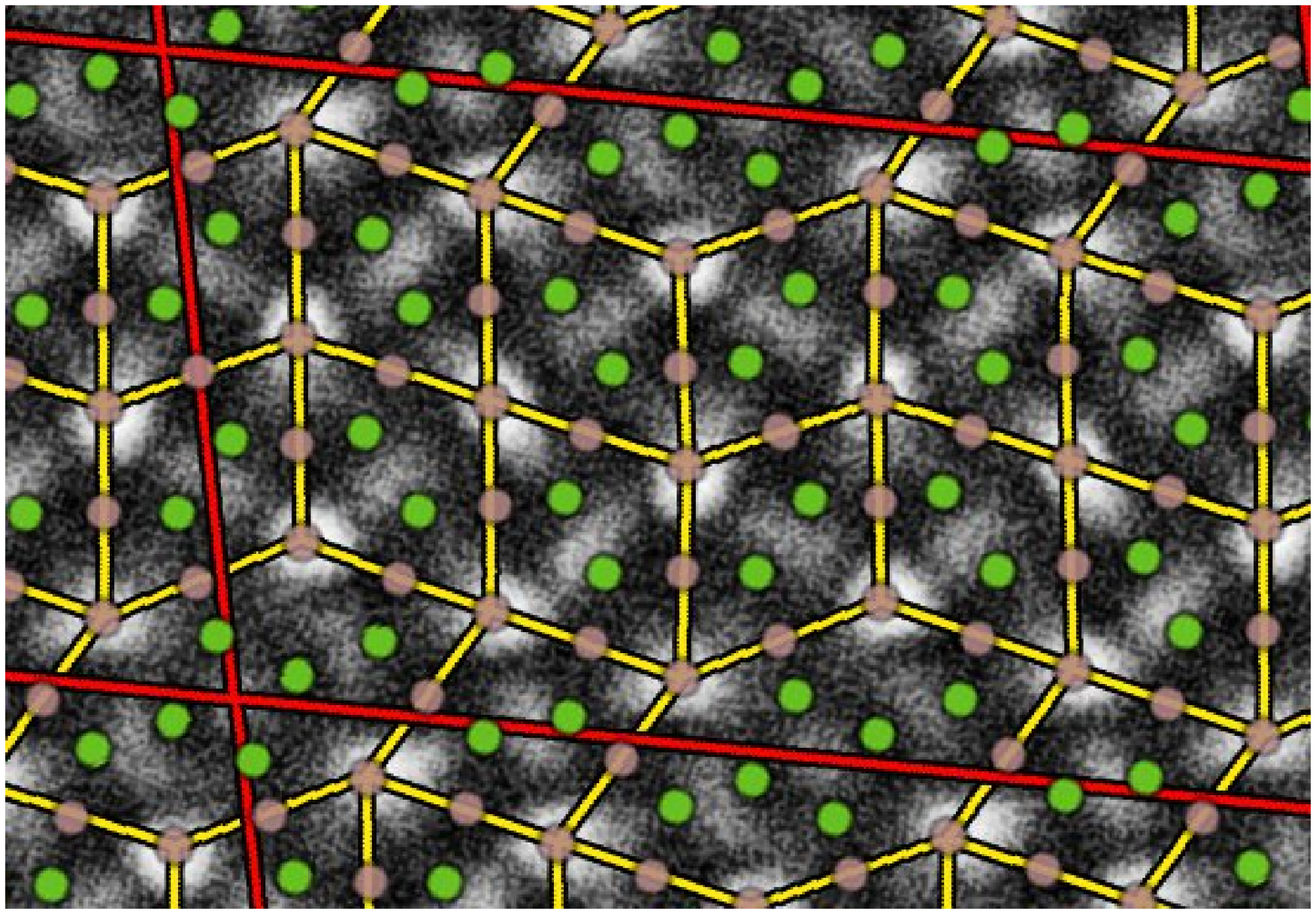}}\hfill\
	\caption{\label{simulations} Multi-slice image simulations of different regions of the precipitate shown in Figure~\protect\ref{fig-hr-1}. 
	Experimental images are shown to the left, with the rhombohedral tiling superimposed.
	Simulations are shown to the right, overlaid with the atomic positions, unit cell boundaries and rhombohedra tiling.
	The simulations shown are for a defocus of -8\,nm and thicknesses of 27.9nm for the Laves phase structures and 27.7\,nm for the monoclinic phase. 
 }
	\end{center}
\end{figure}

The experimentally-obtained images were compared with multi-slice image simulations generated using JEMS \cite{StadelmannJems1987}.  
Figure~\ref{simulations} shows HRTEM images and corresponding simulated images for the three regions of the precipitate in Figure~\ref{fig-hr-1}. 
The experimental images are shown to the left (Figures~\ref{simulations}(a,c,e)) with the corresponding simulation to the right. 
The simulations are for a defocus of 24\,nm and thicknesses of 27.9nm for the Laves phase structures and 27.7\,nm for the monoclinic phase. 
The relevant unit cells, atomic positions, unit  and rhombohedral tiling have been superimposed on the simulations. 
The strongest features in the experimental images (Figures~\ref{simulations}(a,c,e)) are a series of intensity maxima which correspond to the vertices of tiled rhombohedra.	
The three different regions of the precipitate show clearly different arrangements of these sites which are well reproduced in the simulations. 
The simulations (Figures \ref{simulations}(b,d,f)) predict local intensity maxima at Mg sites close to the centres of the rhombohedra  These are observable in some regions of the experimental images, for example to the right in the C14  and monoclinic phase images (Fig. \ref{simulations}(a,e))
In other regions and in the proposed C15 structure, weaker spots occur at many sites midway between the strong intensity sites (i.e. the midpoints of the edges of the rhombohedra, where a Zn atom is located in each structure.)
This suggests that while the rhombohedral sites are well-ordered there may be disorder, perhaps including mixed-occupancy at other sites.
It is also noticeable that the contrast in experimental images changes where the orientation of the rhombohedra in the C14 hexagonal Laves phase and monoclinic structure change.

\begin{figure}
	\begin{center}
		\hfill
		\subfigure[]{\includegraphics[scale=0.25]{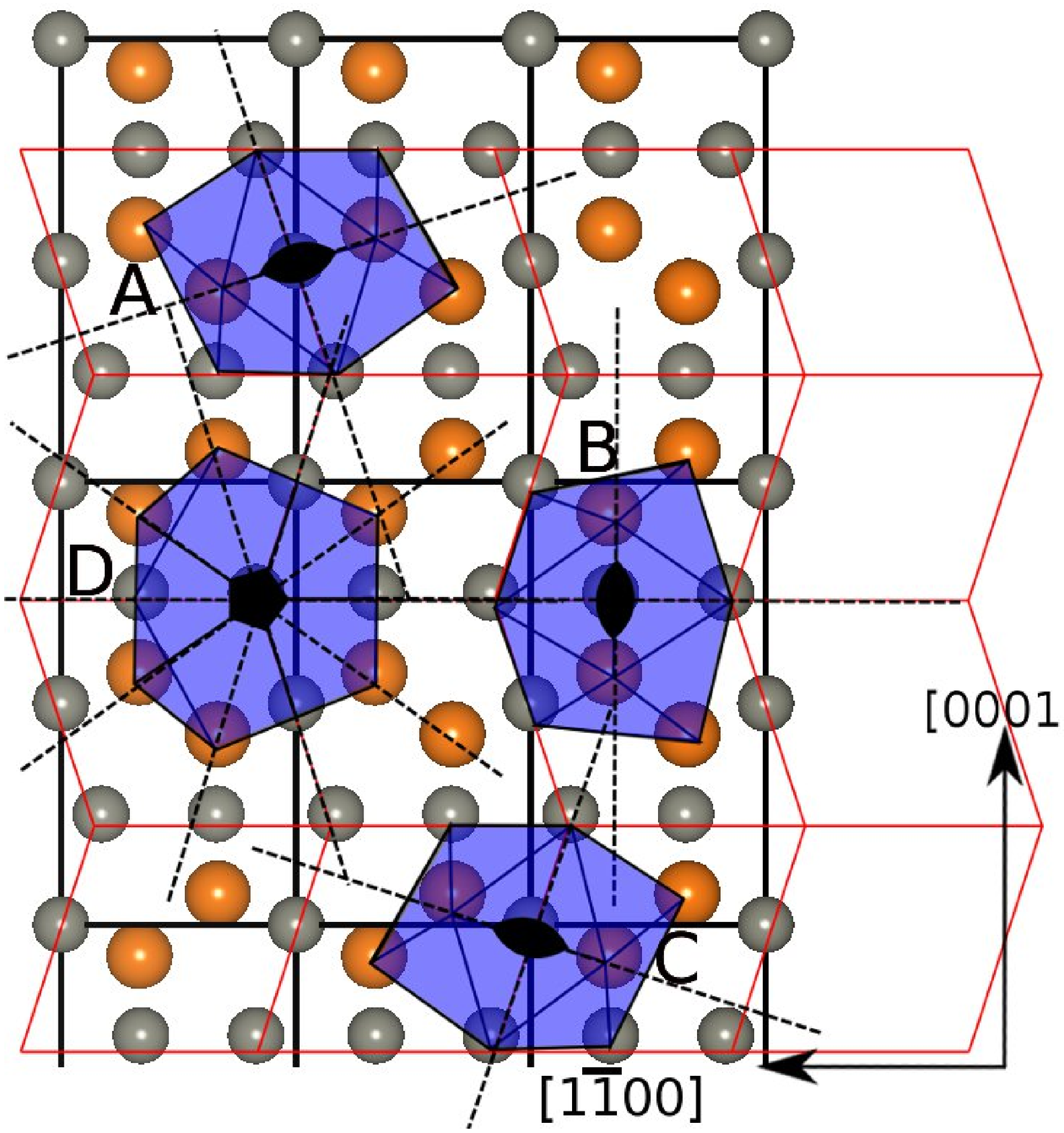}}
		\hfill
		\subfigure[]{\includegraphics[scale=0.25]{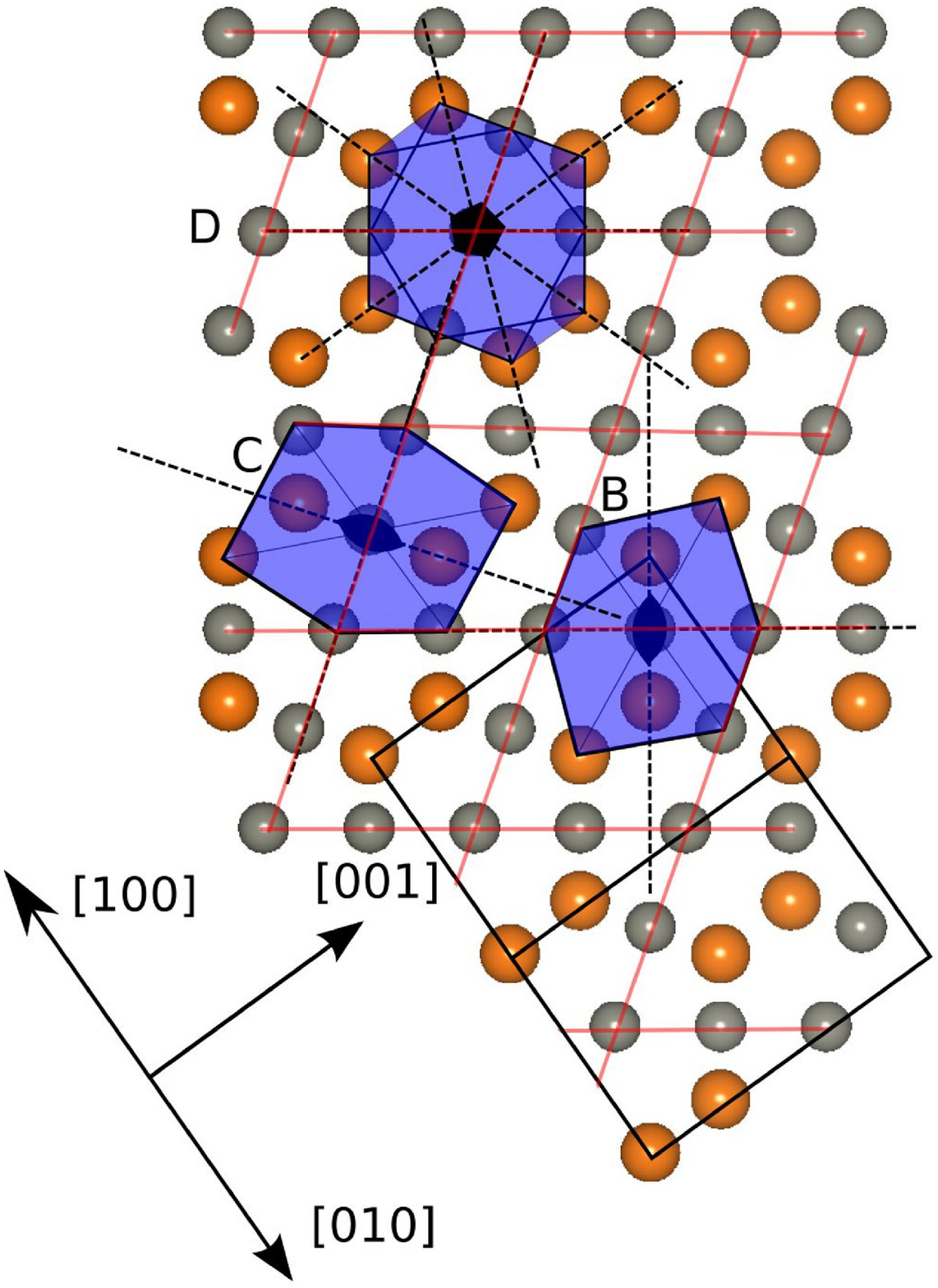}} 
		\hfill\

		\hfill
		\subfigure[]{\includegraphics[scale=0.35]{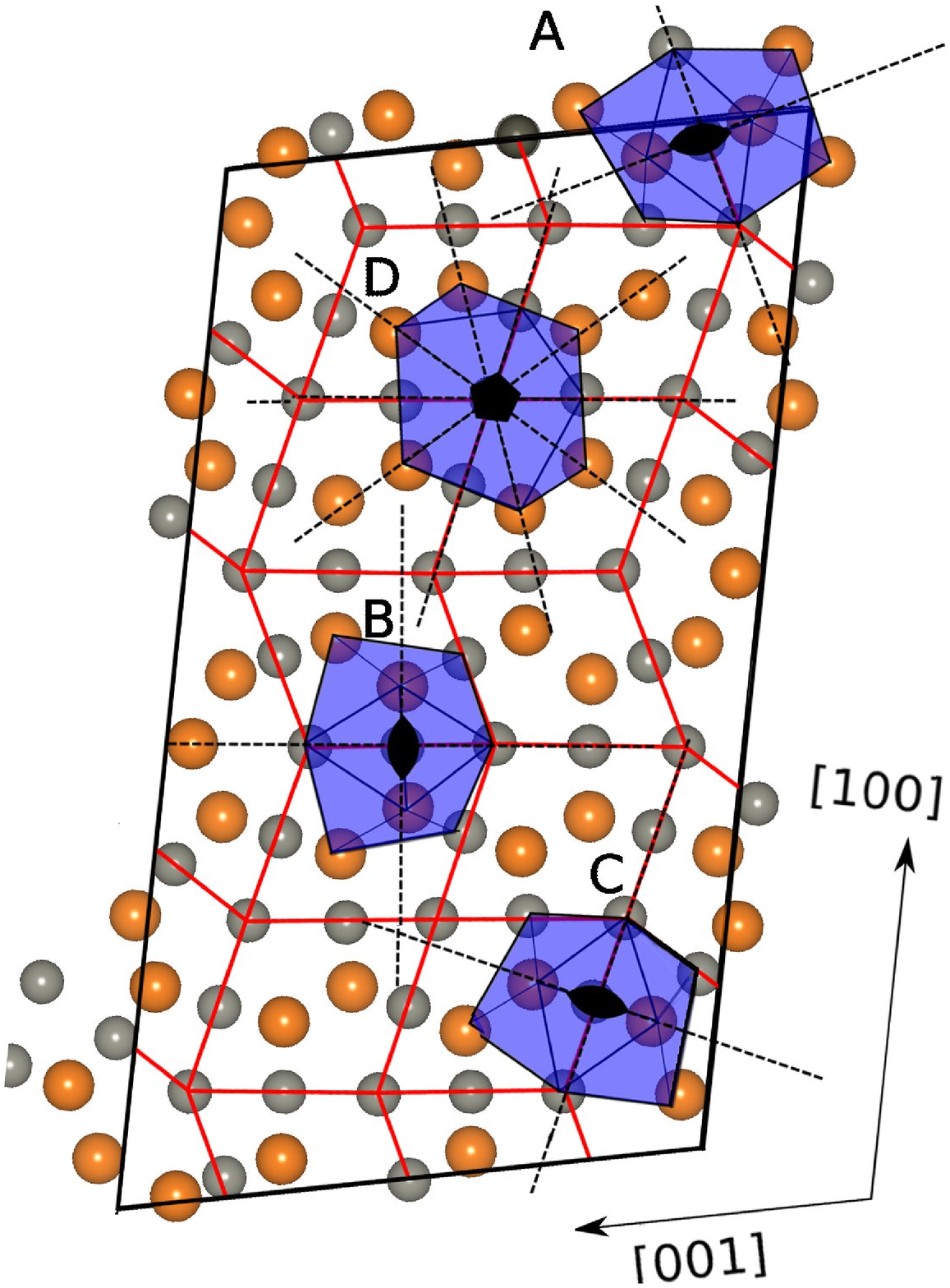}}
		\hfill
		\subfigure[]{\includegraphics[scale=0.35]{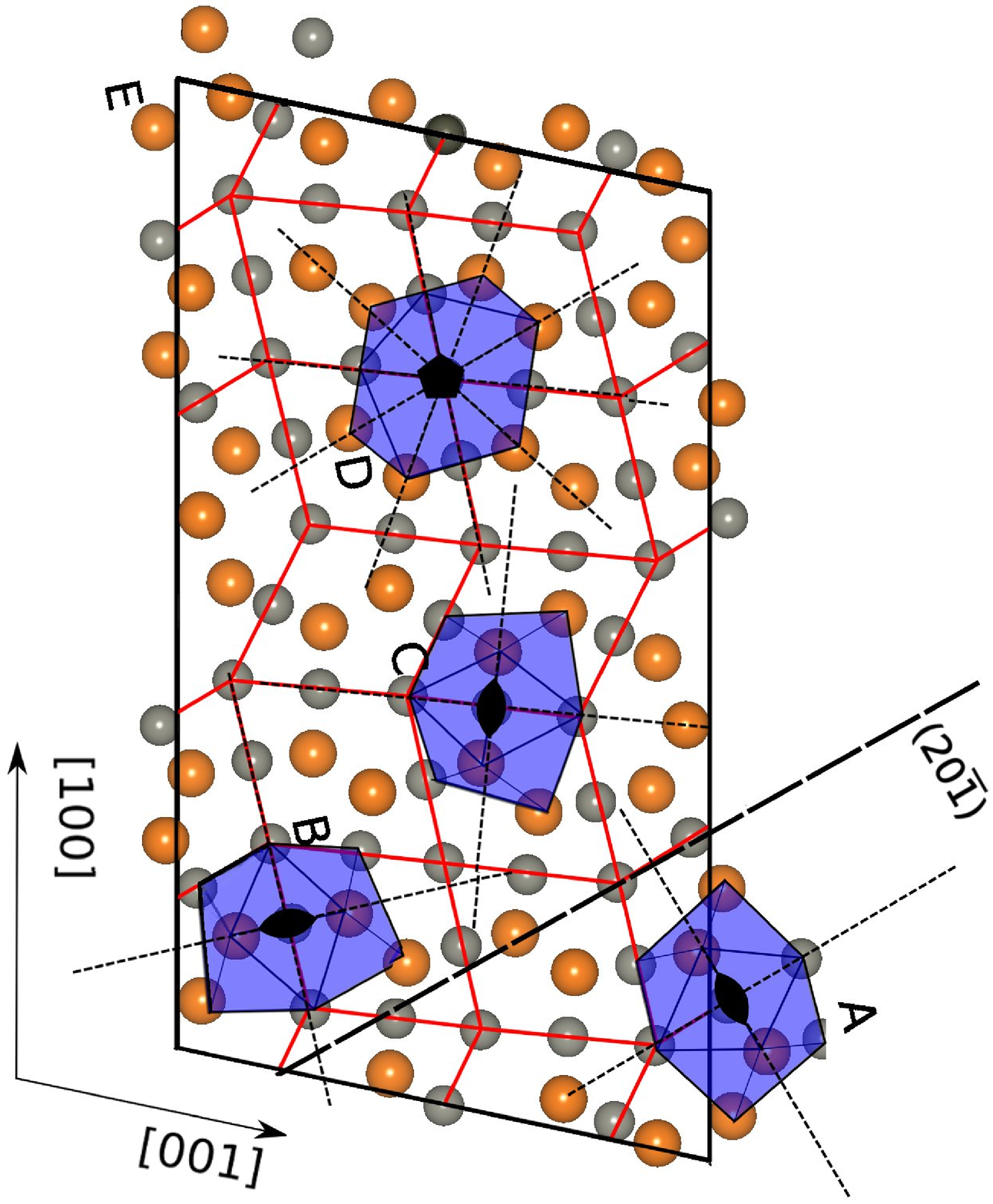}} 
		\hfill\

\caption{Schematics illustrating the orientation of the icosahedral zinc-centred clusters in (a) hexagonal \ce{MgZn2}   (b) C15
 cubic \ce{MgZn2}. and (c,d) two different orientations of the monoclinic \ce{Mg4Zn7} phase.
The icosahedra are viewed with either 5-fold or 2-fold axes, parallel to the viewing direction, which corresponds to 
\( [010]_\ce{Mg4Zn7}\) or \( [11\overline{2}0]_{(hex.)\ce{MgZn2}}\).	
The unit cell boundaries are indicated by solid black lines, and the rhombohedral units by lighter lines.  
\label{fig-icos-5}}

	\end{center}
\end{figure}

Figure~\ref{fig-icos-5}  shows structural models of the (a) hexagonal \ce{MgZn2}   (b) cubic 
 \ce{MgZn2}. and (c,d) the two different orientations of the \ce{Mg4Zn7} phase. present in  the HRTEM image. 
Several, differently oriented icosahedral clusters have been indicated ($A$-$E$) in each figure.
The two Laves phase structures have the orientation relationship;
\( [1\overline{1}00]_{C14} \parallel [110]_{C15}
\);
\(  (0002)_{C14} \parallel (1\overline{1}1)_{C15}
\)
Four differently oriented icosahedral clusters ($A$-$D$) are present in the hexagonal phase, with three identically oriented icosahedra ($B,C,D$) also present in a cubic Laves phase structure. 
For three of these ORs ($A$-$C$), the icosahedron is viewed along a 2-fold (cube direction), with a second 2-fold axis directed along the edge of the rhombohedron. 
These ORs will be of type
\(  [0/0~0/0~/0/2] \parallel [0001]_\mathrm{Mg}\); 
with \(  [0/0~0/2~0/0] \) parallel to  (0002)\(_{C14}\),  \( (20\overline{2}3)_{C14}\) or  \( (\overline{2}023)_{C14}\)
In the fourth OR, marked $D$, the 5-fold axis is parallel to the hexad axis of  Mg and lies in the viewing direction, with 2-fold axis along edges of the rhombohedra and the  
OR will be \(  [1/0~0/1~/0/0] \parallel [0001]_\mathrm{Mg}\). 

Two differently oriented variants of the \ce{Mg4Zn7} structure were present in the HRTEM image and the constituent icosahedra in each phase share common ORs with those in the two Laves phases. 
Figure~\ref{fig-icos-5}(c) shows the monoclinic phase with the [100] direction rotated approximately 6$^o$ clockwise from 
\( [0001]_{C14}\).
In the second variant 	 \( (20\overline{1})_\mathrm{Mg4Zn7}\) is parallel to \( (\overline{2}023)_{C14}\)
In both instances icosahedra clusters ($A$-$D$) adopt identical orientations to those in the hexagonal laves phase. 
For the present orientation of the three phases, there is a very high correspondence between the alignment of the constituent icosahedra, and hence the positions of the peripheral atoms in the zinc-centred clusters.

\section*{Discussion}

Twin boundary and intragranular precipitates in Mg-Zn(-Y) alloys showed the presence of hexagonal \ce{MgZn2} and \ce{Mg4Zn7} phases, with some indications of the presence of a C15 cubic Laves phase structure. 
HRTEM images and image simulations showed that within individual precipitates the local arrangement of icosahedral zinc sites was consistent the structures of these phases. 
The co-existence of hexagonal \ce{MgZn2} and \ce{Mg4Zn7} phases as has recently been shown in ternary Mg-Zn-Y alloys \cite{SinghPhilMag2010, RosaliePhilMag2010}, however binary Mg-Zn alloys had not been examined and it was not possible to preclude yttrium from playing a role in the formation of these dual-phase precipitates. 
On the basis that co-existing (hexagonal)~\ce{MgZn2}-\ce{Mg4Zn7} precipitates are also found in Mg-Zn alloys (as in Figure~\ref{fig-hrtem-1}) it can be argued that this phenomenon is not dependent on the presence of yttrium.

Despite the structural similarities between the i-phase and the \ce{Mg4Zn7} and \ce{MgZn2} phases, at present there is no evidence to suggest that the rod-like or twin-boundary precipitates are ternary in nature.
Similar arrangements of edge-sharing icosahedra were observed in binary and ternary alloys.
Previous studies have not detected yttrium or rare earth elements in the matrix or dendrites \cite{Singh2007,WeiPrec1995} in Mg-Zn-(RE/Y) alloys. 
In addition, Laves phases such as \ce{MgZn2} have been found to exhibit a finite solubility range only where the metallic radii of the elements (\( r_a/r_B\)) 
was close to the ideal value of \( \approx 1.225 \) \cite{ThomaLaves1995}.  
In the case of  \ce{MgZn2}, emprical values give \( r_\mathrm{Mg}/r_\mathrm{Zn} \approx 1.11\), close to the limits found for Laves phases. 
In addition, the ternary elements which were found to be soluble in a Laves phase were those with metallic radii between those of the two major elements.
In the present system, however, yttrium (180\,pm)  is larger in diameter that either Mg (150\,pm) or Zn (135\,pm). 
Instead yttrium, which is only moderately soluble in magnesium, is thought to 
segregate to the grain boundaries in the form of grain boundary precipitates such as the quasi-crystalline \(i\)-phase \((\mathrm{Mg_3YZn_6})\) \cite{Tsai1994}. 

Precipitates in Mg-Zn-Y alloys have been reported to form domains of the cubic and hexagonal \ce{MgZn2} phases \cite{KimLaves2010}.
It was suggested that the transformation between the two phases could be described by a process of synchroshear parallel to the basal plane of the hexagonal phase in a process recognised for other Laves phases \cite{KumarSynchroshear2004}.  
Recent work on twin-boundary precipitates in Mg-Zn-Y alloys has shown that \ce{Mg4Zn7} and hexagonal \ce{MgZn2} phases can co-exist with the interface comprised of a transitional structure similar to \ce{Mg4Zn7}, but with the addition of additional rhombohedral units  \cite{RosaliePhilMag2010}.
Studies of intragranular precipitates showed co-existence of the  \ce{MgZn2} and \ce{Mg4Zn7} phases at a much finer scale, with a largely continuous network of rhombohedra forming either phase \cite{SinghPhilMag2010}. 
In each cases the interfaces could be described by edge-sharing rhombohedra, allowing the two phases to maintain good correspondence between the locations of the zinc-centred icosahedral forming the corners of the rhombohedral arrangement.
Planar defects along which the orientation of the rhombohedra changed marked the boundaries between micro-domains of each phase. 

The cubic and hexagonal Laves phase structures and the monoclinic phase contain similar co-ordination clusters, consisting of 12-coordinate zinc-centred icosahedra. 
In the present work, analysis of the HRTEM images showed that hexagonal \ce{MgZn2} and \ce{Mg4Zn7} domains in twin-boundary and rod-shaped precipitates were aligned so that at least one series of these constituent icosahedra clusters also shared a  common alignment.
This alignment would also be shared by icosahedral clusters in regions adopting a C15 cubic Laves phase structure. 
For an interface formed in such a manner, the peripheral atoms in the icosahedral clusters would share common positions. 
The existence of a network of icosahedrally coordinated zinc atoms (as indicated by the continuous rhombohedral arrangement of these atomic sites) with similarly oriented icosahedral clusters at the interface would permit a high correspondence of atomic sites at a suitably aligned \ce{MgZn2}-\ce{Mg4Zn7} interface. 
This degree of atomic correspondence would allow domains of the two phases to co-exist without large displacements of the peripheral atoms in the clusters.
In essence, the first co-ordination shell remains the same across the interface, with rearrangements occurring at second-nearest neighbour positions. 
This feature may explain the apparently ready co-existence of the Laves and monoclinic  phases.  

\section*{Conclusions}
Both intragranular precipitates and precipitates formed on twin-boundaries in  Mg-Zn and Mg-Zn-Y alloys were investigated via high-resolution transmission electron microscopy.  
Precipitates on twin-boundaries in both binary and ternary alloys were comprised of domains with the structures of the  \ce{Mg4Zn7} and hexagonal \ce{MgZn2} phases.
These phases had a rational orientation relationship with on another and with the magnesium matrix, such that 
\( [0001]_\mathrm{MgZn_2} \parallel  [010]_\ce{Mg4Zn7} \parallel [0001]_\mathrm{Mg}\); 
\( (0\overline{1}10)_\mathrm{MgZn_2} \parallel (20\overline{1})_\ce{Mg4Zn7} \parallel (11\overline{2}0)_\mathrm{Mg}\).
Intragranular, rod-shaped precipitates contained finer domains, with small regions having the structures of the 
\ce{Mg4Zn7} and hexagonal (C14)  \ce{MgZn2} phases, with some suggestion of a cubic C15 \ce{MgZn2} phase. 

An analysis of the twin-boundary and intragranular phases showed that a high proportion of the icosahedral clusters in each phase shared a common alignment with one another in  which either 2-fold or 5-fold axes were parallel to the hexad axis of the magnesium matrix. 
This orientation relationship between the clusters permits a high correspondence between atomic sites at the  \ce{MgZn2}-\ce{Mg4Zn7} interface and may contribute to the ability of the  \ce{Mg4Zn7} and \ce{MgZn2} to co-exist at a nanometre scale.

\section*{Acknowledgements}

One of the authors (JMR) gratefully acknowledges the support of the Japan Society for the Promotion of Science (JSPS) through a JSPS fellowship. 
The authors also thank  Reiko Komatsu, Keiko Sugimoto and Toshiyuki Murao for assistance with sample preparation.

\end{document}